\documentclass{jetpl}

\twocolumn
\lat

\usepackage{graphicx}

\def \be#1\ee {\begin{equation}#1\end{equation}}

\newcommand{\br}{{\bf r}}

\DeclareMathOperator{\str}{str}
\DeclareMathOperator{\Ai}{Ai}

\newcommand{\gap}{E_{\rm g}}
\newcommand{\sg}{s_{\rm g}}
\newcommand{\ETh}{E_{\rm Th}}
\newcommand{\dos}{\left<\rho\right>}

\newcommand{\sx}{\sigma_x}
\newcommand{\sy}{\sigma_y}
\newcommand{\sz}{\sigma_z}
\newcommand{\tx}{\tau_x}

\newcommand{\tz}{\tau_z}

\newcommand{\tf}{\theta_{\rm F}}
\newcommand{\tb}{\theta_{\rm B}}
\newcommand{\kb}{k_{\rm B}}
\newcommand{\chib}{\chi_{\rm B}}

\newcommand{\teps}{\tilde\varepsilon}
\newcommand{\tG}{\tilde G}

\newcommand{\GN}{G_{\rm N}}
\newcommand{\GT}{G_{\rm T}}

\newcommand{\eps}{\varepsilon}

\begin{document}

\title{Density of States in a Mesoscopic SNS Junction}
\rtitle{Density of States in a Mesoscopic SNS Junction}
\sodtitle{Density of States in a Mesoscopic SNS Junction}

\author{P.\,M.\,Ostrovsky\thanks{e-mail: ostrov@itp.ac.ru},
         M.\,A.\,Skvortsov, and M.\,V.\,Feigel'man}
\rauthor{P.\,M.\,Ostrovsky, M.\,A.\,Skvortsov, M.\,V.\,Feigel'man}
\sodauthor{Ostrovsky, Skvortsov, Feigel'man}

\address{Landau Institute, Kosygina 2, Moscow, 117940, Russia}


\abstract{
Semiclassical theory of proximity effect predicts a gap $\gap \sim \hbar D/L^2$
in the excitation spectrum of a long diffusive
superconductor/normal\ch metal/superconductor (SNS) junction. Mesoscopic
fluctuations lead to anomalously localized states in the normal part of the
junction. As a result, a non\d zero, yet
exponentially small, density of states (DOS) appears at energies below $\gap$. In the
framework of the supermatrix nonlinear $\sigma$\d model these prelocalized states are
due to instanton configurations with broken supersymmetry. The exact result
for the DOS near the semiclassical threshold is found provided the dimensionless
conductance of the normal part $\GN$ is large. The case of poorly transparent
interfaces between the normal and superconductive regions is also considered.
In this limit the total number of the subgap states may be large.}

\PACS{73.21.-b, 74.50.+r, 74.80.Fp}

\maketitle

{\bf 1. Introduction.}
It has recently been shown within several different although
related contexts that
the excitation energy spectrum of  Superconductive\ch Normal (SN)
chaotic hybrid structures~\cite{Vavilov,OSF01}
and superconductors with magnetic impurities~\cite{Lamacraft00,Lamacraft01}
does not possess a \emph{hard gap} as predicted by a number of
papers~\cite{Golubov,Zhou,Melsen,Pilgram} using the
semiclassical theory of superconductivity~\cite{Eilen,classics,Usadel}.
With mesoscopic fluctuations taken into account, the phenomenon of
\emph{soft gap} appears: the density of states is nonzero at all energies,
but it decreases exponentially fast  below the semiclassical threshold
$\gap \sim \hbar/\tau_c$, with $\tau_c$ being the characteristic
dwell time in the N region.
In particular, for diffusive systems perfectly connected to a superconductor,
$\gap$ has the order of the Thouless energy $\ETh$ 
in the N region~\cite{Golubov,Zhou}.

The first result in this direction was obtained in Ref.~\cite{Vavilov},
where the subgap DOS in a quantum dot was studied by employing 
the universality hypothesis and the predictions~\cite{Tracy}
of the random matrix theory (RMT)~\cite{Mehta}.
Later on, the tail states in a superconductor with magnetic
impurities were analysed in Refs.~\cite{Lamacraft00,Lamacraft01}
on the basis of the supersymmetric nonlinear $\sigma$\d model 
method~\cite{Efetov} extended to include superconductive
paring~\cite{AST}. 

Fully microscopic approach to the problem of the subgap states in
diffusive NS systems was developed in Ref.~\cite{OSF01} in the framework
of the supersymmetric $\sigma$\d model similar to that employed in
Refs.~\cite{Lamacraft00,Lamacraft01}.
Physically, the low\d lying excitations in SN structures are due to anomalously
localized eigenstates~\cite{MK} in the N region. From the mathematical side,
nonzero DOS at $E < \gap$ comes about when  nontrivial
field configurations \t\ instantons \t\ are taken into account in the
$\sigma$\d model functional integral.  As shown in~\cite{OSF01},
at $E\approx \gap$ there are two different types of instantons,  their
actions being  different  by the factor 2.   The main contribution
to the exponentially small subgap DOS is determined by the
Gaussian integral near the least\d action instanton.

For a planar (quasi\d 1D) SNS junction with ideally transparent SN interfaces
it is given by (provided $\GN^{-2/3} \ll \eps \ll 1$)
\be
\dos =
0.97\delta^{-1}\GN^{-1/2}\eps^{-1/4}\exp\left[-1.93\GN\eps^{3/2}\right],
\label{1inst}
\ee
where
$\eps=(\gap-E)/\gap$,
$\GN=4\pi\nu D L_yL_z/L_x \gg 1$ is the
dimensionless conductance (in units of $e^2/2\pi\hbar$) of the
normal part connecting two superconductors,
$\gap=3.12\ETh$, $\ETh=D/L_x^2$ is the Thouless energy,
and $\delta=(\nu V)^{-1}$ is the mean level spacing.
Here $L_x$ is the thickness of the N region, assumed to be larger than
the superconductive coherence length. It is also assumed that
the lateral dimensions $L_y,L_z$ are not much larger than $L_x$
(otherwise, the instanton solution acquires additional dimension(s)
and the exponent $3/2$ changes, cf.~\cite{OSF01} for details).
The corresponding mean\d field (MF) expression above the gap 
reads~\cite{Zhou}
\be
\dos_{\rm MF} = 3.72\delta^{-1}\sqrt{|\eps|}.
\label{sqrtdos}
\ee

Generally, the functional form of Eqs.~(\ref{1inst}), (\ref{sqrtdos})
is retained, whereas the coefficients are
geometry\d dependent and can be found from the solution of the standard Usadel
equation~\cite{Usadel} for the specific sample geometry. In any case,
the total number of states with energies below $\gap$ is of the order
of one.

In the present Letter we extend our previous results~\cite{OSF01} in
two different directions. Firstly, we derive \emph{exact} expression for
the DOS in the energy region $|\eps| \ll 1$, without using anymore the
inequality $\eps \gg \GN^{-2/3}$. The obtained result interpolates
smoothly between the semiclassical square\d root edge~(\ref{sqrtdos})
and exponential tail~(\ref{1inst}). Secondly, we
consider the same SNS system allowing for non\d ideal transparencies
at the SN interfaces. The result depends upon the relation between
dimensionless interface conductance $\GT$ and normal conductance $\GN$.
As long as $\GT \geq \GN^{1/4}$, all  qualitative features of the
previous solution are retained, but the value of the semiclassical threshold
$\gap$ and numerical coefficients in the expression like (\ref{1inst})
become dependent upon the value of $t \equiv \GT/\GN$. However,
at further decrease of interface transparency, $\GT \ll \GN^{1/4}$,
the DOS behaviour changes dramatically: in the semiclassical region
$E > \gap$ it acquires the inverse\d square\d root singularity,
$\dos_{\rm MF} \sim (-\eps )^{-1/2}$. At smallest $|\eps|$ this singularity
smoothens out and crosses over to an exponentially decaying tail of
low\d energy states. Distinctive feature of this tail, as opposed to the
situations discussed previously, is that the total number of
subgap states becomes large and grows as
$\GT^{-1/2}\GN^{1/8}\gg1$. We coined this situation as ``strong tail'',
and find exponential asymptotics of the DOS in the strong tail region.

{\bf 2. Outline of the method.}
We treat the problem within the supersymmetric formalism.
The derivation of the $\sigma$\d model functional\d integral
representation can be found in Refs.~\cite{Efetov,AST,OSF02}.
The DOS is given by the integral over the supermatrix $Q$:
\begin{gather}
\left<\rho(E,\br)\right> = \frac {\nu}4\Real\int\str(k\Lambda Q(\br))
   e^{-{\cal S}[Q]}{\cal D}Q,
\label{rho}\\
{\cal S}[Q] = \frac{\pi\nu}8\int d\br\;\str\left[
   D(\nabla Q)^2+4iQ(\Lambda E+i\tx\Delta)
\right].
\label{act}
\end{gather}
$Q$ is an $8\times 8$ matrix operating in Nambu, time\d reversal and Bose\ch Fermi
(supersymmetry) spaces. Pauli matrices operating in Nambu and TR spaces are
denoted $\tau_i$ and $\sigma_i$. The matrix $k$ is the third Pauli matrix
in FB space. $\Lambda=\tz\sz$. Integration in~(\ref{rho}) runs over
the manifold $Q^2=1$ with the additional constraint
\be
Q=CQ^TC^T,\qquad
C=-\tx\begin{pmatrix}
i\sy&0\\
0&\sx
\end{pmatrix}_{\rm FB}.
\ee
This manifold is parameterized by 8 commuting and 8 anticommuting variables.
It turns out however that only 4 commuting and 4 anticommuting modes are
relevant in the vicinity of the quasiclassical gap while contributions
from all other modes to the DOS cancel. The detailed discussion of this fact
will be published elsewhere~\cite{OSF02}. The reduced parameterization
for the commuting part of $Q$ in terms of the 4 variables reads~\cite{OSF01}:
\begin{align}
  Q_c^{\rm BB} &= [\sz\cos\kb +
  \tz\sin\kb (\sx\cos\chib + \sy\sin\chib)]\nonumber\\
               &\times[\tz\cos\tb + \sz\tx\sin\tb],
\label{Qc}\\
  Q_c^{\rm FF} &= \tz\sz\cos\tf+\tx\sin\tf.\nonumber
\end{align}

The commuting part of the action (with all Grassmann variables being zero)
is simplified by introducing new variables
$\alpha=(\tb+\kb)/2$, $\beta=(\tb-\kb)/2$. Then the action~(\ref{act}) 
for the normal part ($\Delta=0$) takes the form
\be
  S[\theta_{\rm F},\alpha,\beta] =
    2S_0[\theta_{\rm F}] - S_0[\alpha] - S_0[\beta],
\label{Stab}
\ee
\be
  S_0[\theta] =
    \frac{\pi\nu}4 \int d\br \;
    \left[D(\nabla\theta)^2 + 4iE\cos\theta \right].
\label{S0}
\ee
Variation of this action yields the identical Usadel equations
for $\tf$, $\alpha$, and $\beta$:
\be
D\nabla^2\theta+2iE\sin\theta=0,
\label{usadel}
\ee
with the boundary conditions $\theta(\pm L_x/2) = \pi/2$.
Eq.~(\ref{usadel}) generally possesses two different solutions
$\theta_{1,2} = \pi/2 + i\psi_{1,2}$
which coincide ($\psi_{1,2}({\bf r}) = \psi_0({\bf r})$)
 just at the threshold energy $\gap$, and are close to
each other in the range $|\eps| \ll 1$ we are interested in.
Thus there are 8 possible saddle points for the action~(\ref{Stab})
corresponding to two solutions of the Usadel equation for each variable
$\tf,\alpha,\beta$. Rotation over the angle $\chib \in [0,2\pi)$ connects
some of them and produces the whole degenerate family of saddle points
(see Refs.~\cite{OSF01,OSF02} for details).
In the following, we will need the function $f_0({\bf r})$ which is
the normalized difference $\psi_2({\bf r})-\psi_1({\bf r})$ at
$E\to \gap$; it obeys the linear equation
\be
  D\nabla^2f_0+2\gap\sinh\psi_0\,f_0=0.
\label{f0eq}
\ee

{\bf 3. Exact result for the transparent interface.}
For energies close to $\gap$ we substitute
$\theta=\pi/2+i\psi_0+igf_0$ into~(\ref{S0}), expand it in powers
of $g$ and $\eps$ and integrate over space using~(\ref{f0eq}):
\begin{gather}
  S_0[\theta]=
  S_0[\pi/2+i\psi_0]+\tG\left[-\teps g+\frac{g^3}3\right],
\label{S0cubic}\\
  \tG=\frac{\pi c_2\gap}{2\delta}, \quad
  \teps=\frac{2c_1}{c_2}\eps,
\nonumber
\end{gather}
where we have introduced the constants
$c_n=\int (d\br/V) f_0^{2n-1}\cosh\psi_0$.
For the quasi\d 1D geometry, $c_1=1.15$ and $c_2=0.88$.
To describe deviation of the angles $\alpha$, $\beta$, $\tf$
from $\pi/2+i\psi_0$ we introduce, analogously to $g$,
three parameters $u$, $v$, $w$, respectively.
Grassmann variables are introduced as
$Q=e^{-iW_c/2}e^{-iW_a/2}\Lambda e^{iW_a/2}e^{iW_c/2}$
where $Q_c=e^{-iW_c/2}\Lambda e^{iW_c/2}$ is specified in Eqs.~(\ref{Qc})
and
$$
W_a=\begin{pmatrix}
0&W_a^{\rm FB}\\
i\tx\sx(W_a^{\rm FB})^T\sy\tx&0\\
\end{pmatrix},
$$
as it must satisfy the antiselfconjugate condition $W_a+CW_a^TC^T=0$.
Finally,
$$
W_a^{\rm FB}=\frac{f_0}4\begin{pmatrix}
0&\zeta-\lambda&\zeta+\lambda&0\\
-\zeta+\lambda&0&0&-\zeta-\lambda\\
\xi+\eta&0&0&\xi-\eta\\
0&-\xi-\eta&-\xi+\eta&0\\
\end{pmatrix}.
$$

Expanding the action in $u$, $v$, $w$ and Grassmann variables
leads to
\begin{multline}
  {\cal S} = \tG\biggl[\teps(u+v-2w)-\frac{u^3+v^3-2w^3}3\\
  -\zeta\xi\frac{u+w}4-\lambda\eta\frac{v+w}4\biggr].
\label{Suvw}
\end{multline}
For calculating the DOS we also need an expansion of the pre\d exponential
factor in~(\ref{rho}) as well as the Jacobian $J$ for the parameterization
of the $Q$\d matrix:
\begin{gather*}
  \frac{\nu}4\int d\br\;\str(k\Lambda Q) = -\frac{ic_1}{2\delta}(u+v+2w),
  \\
  J=\frac{8i\tG^2}{\pi}|u-v|.
\end{gather*}
Integrating over Grassmann variables and the cyclic angle $\chib$,
performing a rescaling $(u,v,w)\to(2\tG)^{-1/3}(u,v,w)$ which excludes
$\tG$ from the integrand, and changing the variables to $l=(u+v)/2$,
$m=(u-v)^2/2$, we arrive at the following expression for the integral DOS:
\begin{multline}
\dos = \frac{c_1(2\tG)^{-1/3}}{4\pi\delta}
  \Real\int\limits_0^{\infty}dm\int dl\,dw\; (w+l)\\
  \times(w^2+2lw+l^2-m)\exp
  \left[-\frac{w^3}3+\epsilon w+\frac{l^3}3+ml-\epsilon l\right]\!,
\label{Airyint}
\end{multline}
where we introduced the notation
$
  \epsilon = (2\tG)^{2/3} \teps.
$

At this stage we have to choose the contours of integration over $w$ and $l$.
The usual convergence requirements for the nonexpanded action (\ref{Stab})
enforce the contour for $w$ ($l$) go along the imaginary (real) axis at large
values of $w$ ($l$). However since the main contribution to the DOS comes
from the expression~(\ref{Airyint}) determined by small $w$ and $l$,
these contours should be properly deformed to achieve convergence of
(\ref{Airyint}). The integral~(\ref{Airyint}) converges if the contour
for $l$ runs to infinity in the dark regions in Fig.~1
and otherwise for $w$.
Therefore, we should choose the contour $C_1$ for $w$
(see Fig.~1), whereas for $l$ there are two
possibilities: $C_2$ and $C_3$. The correct choice is dictated
by positivity of the DOS, which imply the contour $C_3$ for $l$.
\begin{figure}
\centering
\includegraphics[width=0.8\columnwidth]{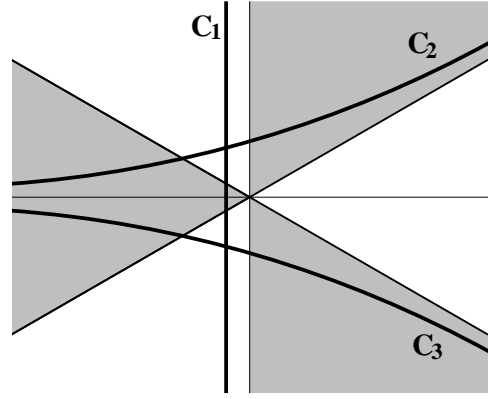}
\caption{Fig.~1. Possible contours for integration over $w$ and $l$. The 
proper choice is $C_1$ for $w$ and $C_3$ for $l$.}
\label{FigContours}
\end{figure}
Integration in~(\ref{Airyint}) is straightforward although rather cumbersome
and leads to the final expression for the DOS:
\begin{multline}
\dos = \frac{2\pi c_1}{(2\tG)^{1/3}\delta} \biggl[
-\epsilon\Ai^2(\epsilon)+[\Ai'(\epsilon)]^2\\
+\frac{\Ai(\epsilon)}2\int_{-\infty}^{\epsilon}dy\;\Ai(y)
\biggr],
\label{Airydos}
\end{multline}
where $\Ai(\epsilon)$ is the Airy function.
Asymptotic behaviour of the calculated DOS at $\epsilon \gg 1$ coincides with
the result (\ref{1inst}) of the single\d instanton approximation~\cite{OSF01},
see Fig.~2.

\begin{figure}
\centering
\includegraphics[width=0.8\columnwidth]{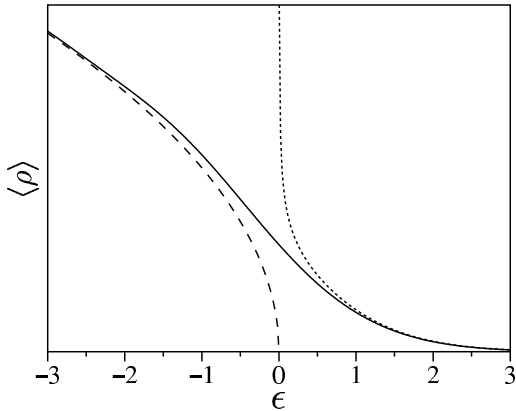}
\caption{Fig.~2. Exact dependence~(\protect\ref{Airydos}) of the DOS on 
the dimensionless energy $\epsilon$ (solid line). Single\d instanton
approximation (\protect\ref{1inst}) (dotted line). Semiclassical result
(\protect\ref{sqrtdos}) for the DOS (dashed line).}
\label{FigAiry}
\end{figure}

The functional dependence~(\ref{Airydos}) coincides with the RMT
prediction for the spectrum edge in the orthogonal ensemble~\cite{Tracy}.
It is not surprising because the random matrix theory
is known to be equivalent to the 0D $\sigma$\d model~\cite{Efetov}.
In our case the problem became effectively 0D when we fixed the coordinate
dependence $f_0(\br)$ for the parameters of $Q$ near $\gap$.

Breaking the time\d reversal symmetry drives the system to the unitary
universality class. The corresponding RMT result~\cite{Tracy} can be obtained
from Eq.~(\ref{Airydos}) by dropping the last integral term.
This result can be easily derived by the $\sigma$\d model analysis
in the following way. Strong magnetic field imposes an additional
constraint on the $Q$\d matrix. As a result the mode associated with
the variable $m$ acquires a mass, so, instead of integrating over it
we set $m=0$. One of the Grassman modes is also frozen out giving the
pre\d exponent $(w+l)^2$ in the integral~(\ref{Airyint}).  Finally, the expression
for DoS coincides with Eq.~(\ref{Airydos}) but without the last term.

{\bf 4. Finite transparency of the NS interface.}
Now we turn to the analysis of the sub\d gap structure of a quasi\d 1D
SNS contact with finite conductance $\GT$ of the NS boundary.
The role of the interface is described by the dimensionless parameter
$t=\GT/\GN$. For $t\gg1$, the interface is transparent and the result
(\ref{Airydos}) apply. In what follows we will consider the case $t\ll1$.
The effect of finite transparency is described by the additional boundary
term~\cite{Efetov} in the action
\be
{\cal S}_{\rm boundary} =
-\frac{\GT}{16}\str\left(Q^LQ^R\right),
\label{Sboundary}
\ee
where $Q^{L,R}$ are the $Q$\d matrices at both sides of the interface.
Eq.~(\ref{Sboundary}) is the the first term in the expansion of the
general boundary action~\cite{Efetov,Lamacraft01,Nazarov} in the
small transparency $\Gamma\ll1$ of conductive channel and leads
to the Kupriyanov\ch Lukichev~\cite{KL} boundary conditions.
In the diffusive regime ($l\ll L_x$) at $t\ll1$, we have $\Gamma\sim tl/L_x$
that justifies the use of Eq.~(\ref{Sboundary}).
The commuting part of the action can still be written in the form~(\ref{Stab})
with the additional term in $S_0$:
\begin{multline}
S_0[\theta] = \frac{\pi\nu L_yL_z}4\int\limits_{-L_x/2}^{L_x/2}dx\;
\left[D(\theta')^2+4iE\cos\theta\right]\\
-\frac{\GT}4\sin\left[\theta\left(\frac{L_x}2\right)\right].
\label{S0t}
\end{multline}

In the limit $t\ll1$, the Usadel equation has almost space
homogeneous solutions, which allows to use an expansion $\psi = A +
B[1-4(x/L_x)^2]$ for them. Substituting this ansatz into the action (\ref{S0t}),
and minimizing over $B$ we obtain the action in terms of $P=e^A$:
\be
  S_0(P) =
  \frac{\GN}8\left[(s-t)P - \frac{2t}P + \frac{t^2}{24}P^2\right] ,
\label{S0P}
\ee
where $s=E/\ETh$. Here we keep only leading terms and substitute all
$s$ except the first one by $t$. After variation we find the cubic saddle
point equation for $P$:
\be
  \frac st = 1 - \frac 2{P^2} - \frac t{12}P.
\label{Pequ}
\ee
The maximum of the RHS achieved at $P_0=(48/t)^{1/3}$ determines
the position of the mean\d field gap: $\sg = t + O(t^{5/3})$
and hence $\gap=\GT\delta/4\pi$.
Depending on the deviation from the threshold,
$\eps=(\gap-E)/\gap=(\sg-s)/\sg$, there are two regimes for Eq.~(\ref{Pequ}).

\emph{Weak tail.}
If $|\eps|\ll t^{2/3}$ the two solutions of~(\ref{Pequ}) are close to each
other and can be seek in the form $P = P_0 + \delta P$. Expanding the
action in powers of $\delta P$ we get
\be
  S_0(P) = S_0(P_0)
    + \frac{\GT}8\left[-\eps\delta P + \frac 2{P_0^4}(\delta P)^3\right].
\label{S0expand}
\ee
This equation closely resembles its counterpart (\ref{S0cubic}) for the
transparent interface. As mentioned in Ref.~\cite{Lamacraft01}, this form
of the expansion of the action over small deviations near the threshold
solution inevitably leads to the instanton action scaling as $\eps^{3/2}$.
In fact, there is full equivalence~\cite{OSF02} between the DOS for the
transparent NS interface given by Eq.~(\ref{Airydos}) and the DOS in the
limit $|\eps|\ll t^{2/3}\ll1$. The latter can be obtained from the former
by redefinition of the constants $c_{1,2}$. For a 1D planar contact they
appear to be $c_1=P_0/2$, $c_2=6/P_0$, $\gap/\ETh=t$.

In particular, above the threshold, at $\eps<0$, one encounters
the mean\d field square\d root singularity
\be
  \langle\rho\rangle_{\rm MF} =
  \frac4{\delta}\frac{6^{1/6}}{t^{2/3}}\sqrt{|\eps|}.
\label{tsqrtdos}
\ee
The instanton action becomes
${\cal S} = S_0(P_1) - S_0(P_2) = (2/3) 6^{1/6}\GN t^{1/3}\eps^{3/2}$,
and the single\d instanton asymptotics of the DOS tail reads
\be
  \dos = \frac 1{\delta}\sqrt{\frac{\pi 6^{1/6}}{2\GN t^{5/3}\sqrt{\eps}}}
  \exp\left(-\frac23 6^{1/6}\GN t^{1/3}\eps^{3/2}\right).
\label{dos32t}
\ee

\emph{Strong tail.}
In the opposite limit, $t^{2/3}\ll|\eps|\ll1$, the difference between the two
solutions for Eq.~(\ref{Pequ}) is large but expansion~(\ref{S0P}) 
is still valid (gradients of $\psi_{1,2}$ are small provided $\eps \ll 1$). 
The roots $P_{1,2}$ can be
found neglecting either the second or the third term in Eq.~(\ref{Pequ}):
$P_1=\sqrt{2/\eps}$,
$P_2=12\eps/t$, with $P_2\gg P_1$. Above the threshold that gives the
\emph{inverse}\d square\d root singularity in the semiclassical DOS:
\be
  \dos_{\rm MF}=\frac 2\nu\Real\int d\br\,\cos\theta
  =\frac1{\delta}\Imag(P-P^{-1})
  =\frac1{\delta}\sqrt{\frac 2{|\eps|}} .
\label{strong-above}
\ee

Below $\gap$ one obtains for the instanton action
${\cal S} = - S_0(P_2) = (3/4)\GN\eps^2$ which determines
the one\d instanton asymptotics of the subgap DOS.
The pre\d exponent can be calculated by generalizing the method
of Ref.~\cite{OSF01}. Introducing the deviation parameter $q$ according to
$\alpha=\pi/2+i\log P_2+iq/\sqrt2$ and expanding the action in powers of $q$
and the corresponding Grassmann pair $\zeta\xi$ we obtain
for the action and the pre\d exponential factor in Eq.~(\ref{rho}):
\begin{gather*}
  {\cal S}=\frac38\GN\eps^2\left[2-q^2+\frac{q\zeta\xi}{2\sqrt2}\right],
  \\
  \frac{\nu}4\int d\br\;\str(k\Lambda Q) =
    -\frac{3i\eps}{t\delta}\left[1-\frac q{\sqrt2}\right].
\end{gather*}
The measure of integration is
${\cal D}Q=2\sqrt{3/t}(2\eps)^{3/4}dq\,d\zeta\,d\xi$.
Inserting these into Eq.~(\ref{rho}) we finally obtain
\be
\dos
=\frac 3{\delta}\sqrt{\frac{\pi}{\GN t^3}\sqrt{\frac{\eps^3}2}}
\exp\left(-\frac 34 \GN\eps^2\right).
\label{strongdos}
\ee

{\bf 5. Discussion.}
We have considered the integral density of states in a coherent
diffusive SNS junction with arbitrary transparency of the SN interface.
For the ideal interface ($\GT\gg\GN$) we managed to go beyond the
single-instanton analysis~\cite{OSF01} and derived the exact
result~(\ref{Airydos}) valid as long as $|E-\gap|\ll\gap$.
This expression uniquely describes the semiclassical square-root DOS
(\ref{sqrtdos}) above the Thouless gap $\gap$, the far subgap tail
(\ref{1inst}), and the crossover region $\eps\sim\GN^{-2/3}$
between the two asymptotics. The functional form of this result
coincides with the prediction of the RMT.

As the SN interface become less transparent, $\GT\ll\GN$, the situation
changes. At $\GT\gg\GN^{1/4}$ these changes are only quantitative:
the position of the quasiclassical gap is shifted to $\gap=(\GT/\GN)\ETh$,
but the DOS both above [Eq.~(\ref{tsqrtdos})]
and below [Eq.~(\ref{dos32t})] the gap has the same dependence
on the deviation $\eps$ from $\gap$,
with the coefficients becoming dependent on $\GT$.
In this limit, the very far part of the tail [at $\eps\gg(\GT/\GN)^{2/3}$]
exhibits another $\eps$-dependence (\ref{strongdos}),
but the corresponding DOS is exponentially small.
Therefore, in the limit $\GT\gg\GN^{1/4}$ the total number of the subgap
states is of the order of 1 and is independent on $\GT$.
We refer to this case as weak tail.

As the interface becomes less transparent, the region of applicability
of the weak tail shrinks and finally disappears at $\GT\sim\GN^{1/4}$.
For even lower $\GT\ll\GN^{1/4}$, the difference between the case of
the transparent interface becomes qualitative: the DOS above $\gap$
acquires an inverse square-root dependence (\ref{strong-above}),
while the subgap DOS follows (\ref{strongdos}).
In this regime the total number of the subgap states is proportional to
$\GT^{-1/2}\GN^{1/8}\gg1$ and grows with decreasing $\GT$
in contrast to all previous cases when this number is of the order of 1.
This indicates that at $\GT\sim\GN^{1/4}$ the universality class of the
problem is changed. At $\GT\ll\GN^{1/4}$ it is no longer equivalent to
the spectral edge of the Wigner-Dyson random matrix ensembles.

The asymptotic results for the DOS above and below the gap,
as well as the width of the fluctuation region near $\gap$
are summarized in Table I for the three regions considered.

\begin{table}
\label{Table}
\begin{center}
\begin{tabular}{|c|c|c|c|}
\hline
& $\eps<0$ & $\eps>0$ & $|\eps|_{\rm fluct}$  \\
\hline
$t\gg1$ & (\ref{sqrtdos}) & (\ref{1inst}) & $\GN^{-2/3}$ \\
\hline
$|\eps|\ll t^{2/3}\ll1$ & (\ref{tsqrtdos}) & (\ref{dos32t}) & $\GN^{-2/3}t^{-2/9}$ \\
\hline
$t^{2/3}\ll|\eps|\ll1$ & (\ref{strong-above}) & (\ref{strongdos}) & $\GN^{-1/2}$ \\
\hline
\end{tabular}
\end{center}
\caption{Table~I. References to the asymptotic formulas for the DOS
above ($\eps<0$) and below ($\eps>0$) the gap, and the width
$|\eps|_{\rm fluct}$ of the fluctuation region for the regimes of
the transparent interface ($t\gg1$), weak ($|\eps|\ll t^{2/3}\ll1$)
and strong ($t^{2/3}\ll|\eps|\ll1$) tails.}
\end{table}

{\bf Acknowledgements.}
This research was supported by the
SCOPES programme of Switzerland, Dutch Organization
for Fundamental Research (NWO), Russian Foundation for Basic Research
under grant 01-02-17759, the programme ``Quantum Macrophysics'' of
the Russian Academy of Sciences, and the Russian Ministry of Science.

\end{document}